\documentclass[12pt]{article}
\usepackage{amssymb,amsmath,epsfig}
\usepackage{cite}
\allowdisplaybreaks

\begin{document}
\title{\bf Stability Analysis of Stellar Radiating Filaments}

\author{Z. Yousaf \thanks{zeeshan.math@pu.edu.pk}, M. Zaeem-ul-Haq Bhatti
\thanks{mzaeem.math@pu.edu.pk} and Ume Farwa \thanks{ume.farwa514@gmail.com}\\
Department of Mathematics, University of the Punjab,\\
Quaid-i-Azam Campus, Lahore-54590, Pakistan.}

\date{}

\maketitle
\begin{abstract}
The aim of this paper is to perform stability analysis of
anisotropic dissipative cylindrical collapsing model in
$f(R,T,R_{\mu\nu} T^{\mu\nu})$ gravity. In this context, the
modified version of hydrodynamical equation is explored by means of
dynamical equations and radial perturbation scheme. We examined the
role of adiabatic index, dissipation as well as the particular
cosmological model on the onset of dynamical instability of the
evolving cylindrical system that was initially in hydrostatic
equilibrium with Newtonian and post Newtonian approximations.
It is pointed out that extra curvature terms of
$f(R,T,R_{\mu\nu}T^{\mu\mu\nu})$ gravity tends to increase the
stability, while that heat radiations push the system to enter into
unstable window. Further, our results reveal the significance of
adiabatic index in the stability analysis of cylindrical celestial
model.
\end{abstract}

\section{Introduction}

Recent cosmic observational data indicate that our universe is
undergoing accelerated expansion phase \cite{ya1}. Further, current
very important results induced by the Planck satellite~\cite{ya2,
Planck:2015xua, Ade:2015lrj}, the BICEP2
experiment~\cite{Ade:2014xna, Ade:2015tva, Array:2015xqh}, and the
Wilkinson Microwave anisotropy probe (WMAP)~\cite{Komatsu:2010fb,
Hinshaw:2012aka} are pointing out the existence of extremely large
amounts of dark energy (DE) and dark matter (DM) with a  specific
proportion to clarify such issue. No experimental evidences at
fundamental level have been found to describe these enigmatic
constituents. There are multiple attempts in literature to discuss
these dark sources elements \cite{v41, b2a, R-DE-MG}.

In the last few decades, number of modified gravity theories have
been explored \cite{z1,z2,z3} to discuss such issues. The simplest
approach to modify GR is $f(R)$ theory in which $R$ is the Ricci
scalar $R$. Initially, Buchdal \cite{6} was among those researchers
who discussed this theory and later on by many relativistic
astrophysicists \cite{7,8,ol6,ol7,ol8}. The representation of
$\Lambda$ governed by cold dark matter $(\Lambda$CDM) model is that
our observed universe is isotropic and flat. The discovery of cosmic
microwave background radiations (CMBR) is good concurrence with that
predicted model. Further, some observational and experimental data
\cite{9} proposed the existence of some energy sources in
anisotropic configuration in our universe. The $f(R)$ gravity theory
has received great attraction to describe the late-time accelerated
speed up of our universe. Nojiri and Odintsov \cite{z5a} explored
some modified gravity models that could assists enough to explore
the DE dominance in our cosmos. Cognola \emph{et al.} \cite{z5b}
classified some observationally consistent $f(R)$ DE models into
four main categories. Durrer and Maartens \cite{10} examined that
$f(R)$ models could lead to valuable insights in structure
formation.

The matter-geometry coupling could be considered as a viable attempt
to explore modified gravity models. Such kind of generalizations
have been introduced by Harko \emph{et al.} \cite{11} and is named
as $f(R,T)$ gravity in which $T$ is the trace of energy momentum
tensor. They obtained the equations of motion for the massive
particles by using variational principle. Afterwards, many people
used this theory to discuss cosmological reconstruction,
thermodynamics, viscous solutions, stability of homogeneous
universe, dynamical instability, \cite{zz1, zz2} etc. The $f(R,T,Q)$
gravity involves usual Lagrangian demonstrating non-minimal
association between geometry and matter, where
$Q=R_{\lambda\sigma}T^{\lambda\sigma}$. The $f(R,T,Q)$ gravity
theory could be convenient for the study of acceleratory conducts in
course of cosmic expansion.

Haghani \emph{et al.} \cite{21} obtained the field equations (FEs)
by using Lagrange multiplier in $f(R,T,Q)$ theory of gravity.
Odintsov and S$\acute{a}$ez- G$\acute{o}$mez \cite{22} studied
$f(R,T,Q)$ gravity with non-minimal association between matter and
gravitational fields and concluded that gravity induced by such
theory contains additional points which could help to recast the
possible cosmological evolution. Elizalde and Vacaru \cite{z20a}
evaluated some exact off-diagonal cosmological models in $f(R,T,Q)$
gravity. Baffou \emph{et al.} \cite{25} used perturbation technique
and performed stability analysis with the help of de-Sitter and
power law models through numerical simulations in $f(R,T,Q)$
gravity. Recently, Yousaf \emph{et al.} \cite{24} investigated
stability of spherical collapse by perturbative analysis of
cosmological model $\alpha R^2+\beta
(R_{\lambda\sigma}T^{\lambda\sigma})$ in the framework of $f(R,T,Q)$
gravity and found that additional curvature invariants considerably
decelerate the collapse rate because of their repulsive behavior.

There is a remarkable role of matter contents in formation and
evolution of stellar system. Initially, Chandrasekhar \cite{14}
investigated the dynamical instability constraints of oscillating
spherically symmetric model by using isotropic fluid contribution. Instability range depends on the critical value $\frac{4}{3}$ for
spherical stars with isotropic fluids as described by Chadrasekhar.
However, with the inclusion of anisotropies due to various pressure
components, the instability range depends on the physical parameters
like energy density, pressure etc. for which the adiabatic index
depends on the energy density and different stresses of the fluid
distribution and contain static terms of the geometry. It is worth
noting that different ranges of instability may lead to different
patterns of evolution of stars. Chan \emph{et al.} \cite{chananiso1,chananiso2}
studied the effects of local anisotropy in the on the existence of relativistic
spherical celestial objects and inferred that anisotropy increases the
instability regimes under certain conditions. Herrera \emph{et al.} \cite{15} analyzed dynamical instability of
massive stars in context of perturbation technique and found that one can obtain instability constraints
independent of adiabatic index by assuming expansion-free fluid within the stellar interior.

In modern cosmology, not only DE but also
gravitational collapse is the remarkable issue. Gravitational
collapse is highly dissipative phenomena for structure formation in
our universe. In GR, gravitational collapse for radiating stars has
been widely studied \cite{16}. De Oliveira \emph{et al.} \cite{17} studied a collapse of the radiating isotropic shear free
star by presenting a relation between pressure and energy density of the fluid and claimed that
their model could be considered as a toy model for the dynamics of Supernova.
Herrera \emph{et al.} \cite{z4,z5} investigated the influences of heat flow
in the evolution of gravitational collapse of stellar interiors.
Chan \emph{et al.} \cite{chanrad} investigated the rate of gravitational collapse of the
spherically symmetric geometry coupled with heat radiating isotropic matter
content and concluded that Newtonian influences coming from dissipation
increases the instability regions. Bonnor \emph{et al.} \cite{18} studied spherical collapse of perfect fluid distribution
along with heat flow by using Darmois matching conditions and
concluded that pressure on the boundary of the respective
collapsing surface is non-vanishing due to heat flux. Cembranos \emph{et al.}
\cite{20} studied gravitational collapse in $f(R)$ gravity and
analyze that this phenomena is salient tool to the constrain models
which shows late time cosmological acceleration. In recent papers,
Yousaf and his coworkers \cite{zza3} investigated the celestial
collapse in modified gravity and discussed the instability of
compact stars by taking into account the particular viable models
with and without perturbation techniques $\Gamma_1$.

In this paper, we examine the influence of non-minimal association
of matter and geometry on the stability of self-gravitating
cylindrical object in the modified GR theory. The paper is designed
as follows. The basic formalism of $f(R,T,Q)$ gravity is given in
section \textbf{2}. In section \textbf{3}, we provide the dynamics
of cylindrical self-gravitating dissipative collapsing model in
which formation of field, dynamical equations and kinematical
quantities are analyzed by perturbation technique. Also the
instability constraints at Newtonian (N) and post-Newtonian (pN)
limits are examined. In the end, we conclude our main findings.

\section{The Formalism of $f(R,T,Q)$ Gravity}

The formalism of $f(R,T,Q)$ gravity is based on the contribution of
non minimal coupling of geometry and matter, where $R$ in the EH
action is replaced with an arbitrary function of $R,~T$ and
$R_{\gamma\delta}T^{\gamma\delta}$. For the formalism of $f(R,T,Q)$
gravity, the modified action along with matter is defined as
\cite{21}
\begin{equation}\label{1}
I_{f(R,T,Q)}=\frac{1} {2}\int d^4x\sqrt{-g}
[f(R,T,R_{\lambda\sigma}T^{\lambda\sigma})+ \textit{L}_m],
\end{equation}
where, $R_{\lambda\sigma}T^{\lambda\sigma}$ and $R,~T$ represent the
contraction of $T^{\lambda\sigma}$ with Ricci tensor, the Ricci
scalar and trace of the energy-momentum tensor, respectively. The
connection of matter lagrangian $\textit{L}_m$ with
$T^{\lambda\sigma}$ is expressed by the following relation
\begin{align}\label{2}
&T_{\lambda\sigma}^{(m)}=-\frac{2}{\sqrt{-g}}\frac{\delta(\sqrt{-g}\textit{L}_m)}
{\delta{g^{\lambda\sigma}}}.
\end{align}
By varying the modified action (\ref{1}), with the metric tensor
$g_{\lambda\sigma}$, we obtained the following the field equations
\begin{align}\nonumber
&-G_{\lambda\sigma}(f_{Q}\textit{L}_m - f_R) - g_{\lambda\sigma}
\left\{\frac{f} {2}-\Box f_R
-\frac{R}{2}f_R-\frac{1}{2}\nabla_\pi\nabla_\rho(f_{Q}T^{\pi\rho})
-\textit{L}_mf_T\right\}\\\nonumber
&+2f_QR_{\pi(\lambda}T_{\sigma)}^{~\pi}+\frac{1}{2}\Box(f_QT_{\lambda\sigma})-\nabla_\pi\nabla_{(\lambda}
[T^\pi_{~\sigma)}f_Q]-2\left(f_Tg^{\pi\rho}+f_QR^{\pi\rho}\right)\frac{\partial^2\textit{L}_m}{\partial
g^{\lambda\sigma}\partial g^{\pi\rho}}\\\label{3} &-
T_{\lambda\sigma}^{(m)}(f_T+\frac{R}{2}f_Q+1)-\nabla_\lambda\nabla_\sigma
f_R=0,
\end{align}
where $G_{\lambda\sigma}, ~\nabla_\pi$ are Einstein tensor and
covariant derivative, respectively. However,
$\Box=g^{\lambda\sigma}\nabla_\lambda\nabla_\sigma$ is the
d'Alembert's operator. It is remarkable to note that by setting
$f(R,T,Q)= f(R, T)$ in Eq.(\ref{3}), one can get the FEs for
$f(R,T)$ gravity \cite{11}. From Eq.(\ref{3}), we obtained the
expression of trace as
\begin{align}\nonumber
&3\Box
f_R+\frac{1}{2}\Box(f_QT)-T(f_T+1)+\nabla_\pi\nabla_\rho(f_QT^{\pi\rho})+
R(f_R-\frac{T}{2}f_Q)\\\nonumber &+(Rf_Q+4f_T)\textit{L}_m
-2f+2R_{\pi\rho}T^{\pi\rho}f_Q -2
\frac{\partial^2\textit{L}_m}{\partial g^{\lambda\sigma}\partial
g^{\pi\rho}}\left(f_Tg^{\pi\rho}+f_QR^{\pi\rho}\right).
\end{align}
In $f(R,T,Q)$ gravity theory, the energy momentum tensor is
non-conserved \cite{25}. This is due to the additional-force acting
on particles that are in motion in gravitation field. However, the
conservation can be guaranteed by using some special aspects. It is
important to note that the matter lagrangian $\textsl{L}_m$ is of
second or superior orders in the metric tensor $g_{\lambda\sigma}$,
which represents that second variation of lagrangian should be non
zero. In spite of that, one can drop this term in case of
relativistic frame which are connected by a specific matter
contents. For perfect fluid, the matter lagrangian has no
distinction and its second variation was taken to be negligible
\cite{11}. Equation (\ref{3}) can be rewritten as
\begin{align}\label{4}
&R_{\lambda\sigma}-\frac{R}{2}
g_{\lambda\sigma}=G_{\lambda\sigma}=\overset{~~\textrm{eff}}{{T}_{\lambda\sigma}},
\end{align}
where the expression of ${{T}_{\lambda\sigma}}^{\textrm{eff}}$ is
\begin{align}\nonumber
{{T}_{\lambda\sigma}}^{\textrm{eff}}&=\frac{1}{(f_R-f_Q\textit{L}_m)}\left
[(f_T+\frac{1}{2}Rf_Q+1)T^{(m)}_{\lambda\sigma}+
\left\{\frac{R}{2}\left(\frac{f}{R}-f_R\right)-\textit{L}_mf_T-\frac{1}{2}\right.\right.\\\nonumber
&\left.\left.\times\nabla_{\pi}\nabla_{\rho}(f_QT^{\pi\rho})\right\}g_{\lambda\sigma}
-\frac{1}{2}\Box(f_QT_{\lambda\sigma})
-(g_{\lambda\sigma}\Box-\nabla_\lambda\nabla_\sigma)f_R-2f_QR_{\pi(\lambda}T^\pi_{~\sigma)}\right.\\\nonumber
&\left.+\nabla_\pi\nabla_{(\lambda}
[T^\pi_{~\sigma)}f_Q]+2\left(f_QR^{\pi\rho}+f_Tg^{\pi\rho}\right)\frac{\partial^2\textit{L}_m}{\partial
g^{\lambda\sigma}\partial g^{\pi\rho}}\right].
\end{align}
It is noted that manifestation of lagrangian density depends upon
nature of matter content, which is not unique \cite{11}. In this
paper, we would take $\textit{L}_m=-\mu$, where $\mu$ is the fluid
energy density, which is independent of respective metric tensor.
Therefore, in above equation last term will vanish.

\section{Dissipative Matter Distribution and Cylindrical Field Equations}

We take a cylindrical self-gravitating system filled with locally
anisotropic and radiating fluid configurations. We assume a $3D$
timelike surface denoted by $\Sigma$ that has differentiated our
manifold $\mathcal{W}$, into two exterior and interior portions. The
exterior region is symbolized with $+$ sign, while $-$ sign
describes configuration of interior one. For $\mathcal{W}^-$
manifold, we consider the configuration of relativistic stellar
system with cylindrically symmetric spacetime \cite{her1}
\begin{equation}\label{5}
ds^2_-=-A^2(t,r)(dt^{2}-dr^{2})+B^2(t,r)dz^{2}+C^2(t,r)d\phi^{2}.
\end{equation}
For the representation of cylindrical symmetry, the following ranges
are imposed on the coordinates
$$-\infty\leq t\leq\infty, ~~ 0\leq r, ~~ -\infty<z<\infty, ~~
0\leq\phi\leq2\pi.$$ We number the respective coordinates $x^0=t$,
$x^1=r$, $x^2=z$ and $x^3=\phi$. The spacetime for $\mathcal{V}^+$
is \cite{v34}
\begin{equation}\label{6}
ds^2_+=-e^{2(\gamma-\upsilon)}(d\nu^2-d\rho^2)+e^{-2\upsilon}\rho^2d\phi^2+e^{2\upsilon}d{z}^2,
\end{equation}
where $\gamma$ and $\upsilon$ are the functions of $\nu$ and $\rho$,
while the coordinates are numbered as
$x^{\beta}=(\nu,~\rho,~\phi,~z)$. The corresponding vacuum field
equations provide
\begin{align}\nonumber
&\rho(\upsilon_\nu^2+\upsilon_\rho^2)=\frac{\tilde{f}-\tilde{R}\tilde{f}_R}{2\tilde{f}_R}e^{2(\gamma-\upsilon)},\\\nonumber
&2\rho\upsilon_\nu\upsilon_\rho=\gamma_\nu,\\\nonumber
&\upsilon_{\nu\nu}-\frac{\upsilon_{\rho}}{\rho}-\upsilon_{\rho\rho}=\frac{e^{2(\gamma-\upsilon)}}{4\rho}
\left(\frac{\tilde{f}-\tilde{R}\tilde{f}_R}{\tilde{f}_R}\right)\left\{\rho
e^{-4\upsilon}+\frac{e^{2\gamma}}{\rho}\right\},
\end{align}
where subscripts $\nu$ and $\rho$ indicate partial differentiations with
respect to $\nu$ and $\rho$, respectively and tilde shows that
the quantities are evaluated with constant values of $T,~R$ and $Q$.
These equations suggest the existence of gravitational field in $f(R,T,Q)$ gravity.
Now, we express the matter contents within the interior spacetime with
\begin{equation}\label{7}
T_{\lambda\sigma}=(P_{r}+\mu)V_{\lambda}V_{\sigma}+P_{r}g_{\lambda\sigma}
+q(L_\lambda V_\sigma+L_\sigma V_\lambda)-K_\lambda
K_\sigma(P_{r}-P_{\phi})-S_{\lambda}S_{\sigma}(P_{r}-P_{z}),
\end{equation}
where $q,~P_{\phi},~P_z,~P_r$ are heat flux and various stress
components, respectively. We choose the fluid to be comoving in
given coordinate system. Then four vectors are defined as
\begin{equation}\label{8}
V_{\lambda}=-A\delta^{0}_{\lambda} ,
K_{\lambda}=C\delta^{3}_{\lambda},
~L_{\lambda}=A\delta^{1}_{\lambda}~\textrm{and}  ~
S_{\lambda}=B\delta^{2}_{\lambda}.
\end{equation}
These vectors obey the following relations
\begin{equation}\label{9}
V^{\lambda}V_{\lambda}=-1, ~
K^{\lambda}K{\lambda}=1=S^{\lambda}S_{\lambda}, ~~
V^{\lambda}K_{\lambda}=V^{\lambda}S_{\lambda}=K^{\lambda}S_{\lambda}=0.
\end{equation}
In order to understand the physical behavior of dynamical system,
the kinematical quantities have great significance. The expansion
scalar, ($\Theta=V^{\lambda}_{~~;\lambda}$), for our cylindrical
geometry leads to
\begin{equation}\label{10}
\Theta=\frac{1}{A}\left(\frac{\dot{A}}{A}+\frac{\dot{B}}{B}+\frac{\dot{C}}{C}\right),
\end{equation}
where the overdot represents the time derivations. The shear tensor
$\sigma_{\lambda\sigma}$ is
\begin{align*}
&\sigma_{\lambda\sigma}=V_{(\lambda;\sigma)}+a_{(\lambda}V_{\sigma)}
-\frac{1}{3}\Theta h_{\lambda\sigma},
\end{align*}
which can also be expressed as
\begin{equation}\nonumber
\sigma_{\lambda\sigma}=\sigma_s\left(S_\lambda
S_\sigma-\frac{h_{\lambda\sigma}}{3}\right)+\sigma_k\left(K_\lambda
K_\sigma-\frac{h_{\lambda\sigma}}{3}\right).
\end{equation}
with
\begin{equation}\label{11}
\sigma_s=-\frac{1}{A}\left(\frac{\dot{A}}{A}-\frac{\dot{B}}{B}\right),
~
\sigma_k=-\frac{1}{A}\left(\frac{\dot{A}}{A}-\frac{\dot{C}}{C}\right).
\end{equation}

In gravitational physics, matter and energy configurations are
defined with the help of the energy-momentum tensor
$T_{\alpha\beta}$. It is no more universal depending upon particular
type of matter and interactions which involve in stellar model. As
the cosmos is composed of large number of various matter fields, it
would be much complicated to signify exact $T_{\alpha\beta}$ even if
someone knows the influences of each field and governing dynamical
equations. In modified gravity like $f(R,T,Q)$ theory, it is
convenient to consider $\overset{~~\textrm{eff}}{{T}_{\alpha\beta}}$
as an effective energy momentum tensor that may represent a
realistic matter source. Then, each non-zero component of
$\overset{~~\textrm{eff}}{{T}_{\alpha\beta}}$ (in our paper) would correspond to
dynamical variables with some physical origins. The 00, 01, 11, 22
and 33 would correspond to energy density, heat flux, pressures
along radial, $z$ and $\phi$ directions, respectively along with the
corresponding dark source terms coming from the non-minimal matter
geometry coupling. The $f(R,T,Q)$ field equations (\ref{4}) for the
cylindrical line element coupled with dissipative anisotropic matter
distribution provide following components
\begin{align}\label{12}
\overset{\textrm{eff}}{\mu}&=\frac{1}{f_R-f_Q\mathcal{L}_M}
\left[\mathcal{L}_Mf_T-\frac{1}{2}(f-Rf_R)+\mu\chi_1+\dot{\mu}\chi_2
+\frac{\ddot{\mu}}{2A^2}f_Q+\frac{\mu''}{2A^2}f_Q\right.\\\nonumber
&\left.+\mu'\chi_3+\frac{P''_r}{2A^2}f_Q+P_r\chi_4+P'_r\left\{\frac{f_Q'}{A^2}-\frac{5A'}{2A^3}f_Q\right\}
-\frac{f_Q}{2A^2B}(\dot{P_z}\dot{B}+P'_zB')\right.\\\nonumber
&\left.-\dot{P_r}\frac{\dot{A}}{A^3}f_Q+P_z\chi_5+P_\phi\chi_6-\frac{f_Q}{2A^2}\left(\dot{P_\phi}\frac{\dot{C}}{C}
-{P_\phi'}\frac{C'}{C}\right)-\frac{\dot{f_R}}{A^2}\left(\frac{\dot{A}}{A}+
\frac{\dot{B}}{B}+\frac{\dot{C}}{C}\right)\right.\\\nonumber
&\left.-\frac{{f_R'}}{A^2}\left(\frac{{A'}}{A}-
\frac{{B'}}{B}-\frac{{C'}}{C}\right)+\frac{f_R''}{A^2}\right],\\\nonumber
\overset{\textrm{eff}}{P_r}&=\frac{1}{f_R-f_Q\mathcal{L}_M}\left[\frac{1}{2}(f-Rf_R)
-\mathcal{L}_Mf_T+\frac{\ddot{f_R}}{A^2}\psi_1+\dot{\mu}
\left(\frac{5\dot{A}}{2A^3}f_Q-\frac{\dot{f_Q}}{A^2}\right)\right.\\\nonumber
&\left.-f_R'\psi_2+P_r\chi_7-\frac{\ddot{\mu}}{2A^2}f_Q+\mu\chi_8+\frac{\mu'A'}{2A^3}f_Q
-\frac{\ddot{P_r}}{2A^2}f_Q+P_z\chi_9+P_\phi\chi_{10}\right.\\\nonumber
&\left.+\frac{f_Q}{2A^2}\left\{\dot{P_z}\frac{\dot{B}}{B}-P_z'\frac{B'}{B}
+\dot{P_\phi}\frac{\dot{C}}{C}-P_\phi'\frac{C'}{C}\right\}+\dot{P_r}\chi_{11}
+P_r'\chi_{12}\right],\\\label{13}
\overset{\textrm{eff}}{P_z}&=\frac{1}{f_R-f_Q\mathcal{L}_M}\left[\frac{1}{2}(f-Rf_R)-\mathcal{L}_Mf_T+\dot{\mu}
\chi_{14}+\mu\chi_{13}-\frac{\ddot{\mu}}{2A^2}f_Q+\frac{\mu'A'}{2A^3}f_Q\right.\\\nonumber
&\left.+P_r\chi_{15}+\frac{\dot{A}\dot{P_r}}{2A^3}f_Q+P_r'\left(\frac{5A'}{2A^3}f_Q-\frac{f'_Q}{A^2}\right)
-\frac{P''_r}{2A^2}f_Q+P_z\chi_{16}+P_z'\chi_{17}-\dot{P_z}\right.\\\label{14}
&\left.\times\chi_{18}-\frac{\ddot{P_z}}{2A^2}f_Q
+\frac{P''_z}{2A^2}f_Q+P_{\phi}\chi_{19}+\frac{f_Q}{2A^2}\left(\dot{P_{\phi}}\frac{\dot{C}}{C}
-P_{\phi}'\frac{C'}{C}\right)+\psi_3\right],\\\nonumber
\overset{\textrm{eff}}{P_\phi}&=\frac{1}{f_R-f_Q\mathcal{L}_M}\left[\frac{1}{2}(f-Rf_R)-\mathcal{L}_Mf_T+\dot{\mu}
\chi_{14}+\mu\chi_{13}-\frac{\ddot{\mu}}{2A^2}f_Q+\frac{\mu'A'}{2A^3}f_Q\right.\\\nonumber
&\left.+P_r\chi_{15}+\frac{\dot{A}\dot{P_r}}{2A^3}f_Q+P_r'\left(\frac{5A'}{2A^3}f_Q-\frac{f'_Q}{A^2}\right)
-\frac{P''_r}{2A^2}f_Q+P_z\chi_{20}+P_\phi'\chi_{23}+\dot{P_\phi}\right.\\\label{15}
&\left.\times\chi_{22}-\frac{\ddot{P_\phi}}{2A^2}f_Q
+\frac{P''_\phi}{2A^2}f_Q+P_{\phi}\chi_{21}+\frac{f_Q}{2A^2}\left(\dot{P_{z}}\frac{\dot{B}}{B}
-P_{z}'\frac{B'}{B}\right)+\psi_4\right],\\\nonumber
\overset{\textrm{eff}}{q}&=\frac{1}{f_R-f_Q\mathcal{L}_M}\left[q''\frac{f_Q}{2A^2}
-\ddot{q}\frac{f_Q}{2A^2}-\dot{q}\left\{\frac{f_Q}{2A^2}\left(
\frac{\dot{B}}{B}+\frac{\dot{C}}{C}+\frac{4\dot{A}}{A}\right)+\frac{\dot{f_Q}}{A^2}\right\}+q\right.\\\nonumber
&\left.\times\frac{f_Q}{A^2}\left\{\frac{A'^2}{A^2}-\frac{\dot{A}^2}{A^2}
-\frac{\ddot{A}}{A}+\frac{A''}{A}-\frac{\dot{A}}{A}\left(\frac{\dot{B}}{B}+\frac{\dot{C}}{C}\right)
+\frac{A'}{A}\left(\frac{B'}{B}+\frac{C'}{C}\right)-\frac{\dot{f_Q}}{A^2}\right.\right.\\\nonumber
&\left.\left.\times\left(\frac{\dot{A}}{A}+\frac{\dot{B}}{2B}+\frac{\dot{C}}{2C}\right)+\frac{f_Q'}{2A^2}
\left(\frac{B'}{B}+\frac{C'}{C}+\frac{4A'}{A}\right)-\frac{\ddot{f_Q}}{2A^2}-\frac{f_Q''}{2A^2}
+\left(1\right.\right.\right.\\\nonumber
&\left.\left.\left.+f_T-\frac{3}{2}Rf_Q\right)\right\}+q'\left\{\frac{f_Q''}{A^2}+\frac{f_Q}{A^2}
\left(\frac{B'}{B}+\frac{C'}{2C}+\frac{2A'}{A}\right)-\frac{\dot{f_R}'}{A^2}+\frac{1}{A^3}\right.\right.\\\label{16}
&\left.\left.\times(\dot{A} f'_R+A'\dot{f_R})\right\}\right]
\end{align}
The quantities $\chi_i$'s containing the combinations of metric
coefficients and their derivatives which are given in Appendix
\textbf{A}. In above mentioned equations prime stands for
$\frac{\partial}{\partial r}$ operator. The value of the Ricci
scalar is
\begin{align}\nonumber
R&=\frac{2}{A^2}\left[\left(\frac{\ddot{A}}{A}+\frac{\ddot{B}}{B}+\frac{\ddot{C}}{C}\right)
-\left(\frac{A''}{A}+\frac{B''}{B}+\frac{C''}{C}\right)
+\frac{1}{A^2}(A'^2-\dot{A}^2)\right.\\\label{17}
&\left.+\frac{1}{BC}(\dot{B}\dot{C} -B'C')\right].
\end{align}

\section{Dynamics}

Filaments are the features of interstellar medium which appear in a
variety of astronomical scenarios and is linked with cosmic web
where it acts like a bridge associating different dense regions of
galaxies. In the study of interstellar and galactic structures,
cylindrically symmetric geometries are taken as an ideal case in
literature to represent filamentary structure of the universe. In
this section, we analyze stable regions of stellar filaments with
anisotropic radiating relativistic interiors in $f(R,T,Q)$ gravity.
In this framework, we shall calculate the hydrodynamical equation
for the cylindrical system, that was initially at hydrostatic
equilibrium, but after some time it will undergo oscillations. We
would also compute unstable epochs of the perturbed anisotropic
cylindrical object.

\subsection{Hydrodynamics}

Dynamics actually deals the motion of objects under action of
forces. It is hypothesized that the dynamics of galaxies are
resulted by massive and invisible matter named as dark matter. The
current cosmic circumstances of matter suggested that many stellar
systems are in state of higher dimensional epoch, where it is rather
difficult to examine the dynamical structure of these objects. To
overcome this problem, one may need to develop a relativistic
mathematical model by considering some unvarying suppositions.
Obviously, invisible precise factors of higher dimensional
gravitational collapse would not be completely underlined through
these results. However, some convenient perceptions could be
attained due to these solutions. In this scenario, the fluid
dynamics could help to set out the methodology to investigate the
development of stellar objects. Here, we deal with the dynamics of
cylindrical stellar collapse by means of dynamical equations, The
well-consistent collapsing model is considered in this theory. In
$f(R,T,Q)$ gravity, the expression of covariant derivative of
effective energy momentum tensor becomes
\begin{align}\label{18}
\nabla^\lambda
T_{\lambda\sigma}&=\frac{2}{Rf_Q+2f_T+1}\left[\nabla_\sigma(\textit{L}_mf_T)
+\nabla_\sigma(f_QR^{\pi\lambda}T_{\pi\sigma})-\frac{1}{2}(f_Tg_{\pi\rho}+f_QR_{\pi\rho})\right.\\\nonumber
&\times\left.\nabla_\sigma
T^{\pi\rho}-G_{\lambda\sigma}\nabla^\lambda(f_Q\textit{L}_m)\right],
\end{align}
which yields two equation of motion for our observed cylindrical
system. Considering $G^{\lambda\sigma}_{~;\sigma}=0$, by using
Eqs.(\ref{12})-(\ref{16}) with $\lambda=0,~1$ this relation gives
\begin{align}\nonumber
&D_t\overset{~~\textrm{eff}}{\mu}+\Theta\left[\overset{~~\textrm{eff}}{\mu}
+\frac{1}{3}\left(\overset{~~\textrm{eff}}{P_{r}}+\overset{~~\textrm{eff}}{P_{z}}
+\overset{~~\textrm{eff}}{P_{\phi}}\right)\right]+\nabla\overset{~~\textrm{eff}}{q}+\frac{1}{3}\left(\overset{~~\textrm{eff}}{P_{z}}
-\overset{~~\textrm{eff}}{P_{r}}\right)(2\sigma_s-\sigma_k)\\\label{19}
&+\frac{1}{3}\left(\overset{~~\textrm{eff}}{P_{\phi}}
-\overset{~~\textrm{eff}}{P_{r}}\right)(2\sigma_k-\sigma_s)+\overset{~~\textrm{eff}}{q}
\left[2a+\frac{1}{A}\left(\frac{B'}{B}+\frac{C'}{C}\right)\right]+Z_1=0,\\\nonumber
&\nabla
\overset{~~\textrm{eff}}{P_{r}}-\frac{1}{A}\left[\left(\overset{~~\textrm{eff}}{P_{z}}
-\overset{~~\textrm{eff}}{P_{r}}\right)\frac{B'}{B}+D_t\overset{~~\textrm{eff}}{q}+\left(\overset{~~\textrm{eff}}{P_{\phi}}
-\overset{~~\textrm{eff}}{P_{r}}\right)\frac{C'}{C}\right]+\left(\overset{~~\textrm{eff}}{\mu}
+\overset{~~\textrm{eff}}{P_{r}}\right)\\\label{20} &\times
a-\frac{1}{3}(\sigma_s-4\Theta+\sigma_k)\overset{~~\textrm{eff}}{q}+Z_2=0,
\end{align}
where superscript $`\textrm{eff}'$ indicates the presence of
$f(R,T,Q)$ terms in the matter variables. These equations of motion
are utilized to examine the properties of collapsing model. After
Chandrasekhar's \cite{14} work on dynamical instability, a lot of
work in this direction has been done on imperfect matter
distribution. Here, we will explain the dynamical instability of
anisotropic and dissipative relativistic cylindrical geometry by
using particular $f(R,T,Q)$ model \cite{27}.
\begin{equation}\label{21}
f(R,T,Q)=\beta R(1+{\alpha}Q),
\end{equation}
where $\alpha, \beta$ are constants. For $\alpha\neq0,~\beta\neq0$,
the inactive conversion of field is redefined  which makes the
respective model physically well-consistent. Here, we are interested
to deal with the model for $\beta=1$ to introduce the consequences
of cosmic expansion including $Q$ parametric quantity. Yousaf \emph{et
al.} \cite{24} considered $f(R,T,Q)=\alpha R^n+\beta Q^m$ model with $n=2,~m=1$ and discussed the
stability of compact stars in anisotropic spherical configuration by
taking $\beta>0$ along with $\alpha=\frac{1}{6M^2}$.

For uninterrupted connections between Eqs.(\ref{5}) and (\ref{6})
over $\Sigma$, we shall make use of Darmois \cite{v36} as well as
Senovilla \cite{sano1} continuous matching conditions. We consider
timelike hypersurface for which, we impose $r=$constant in
Eq.(\ref{5}) and $\rho(\nu)$ in Eq.(\ref{6}). In this context, the
first fundamental form on $\Sigma$, provide
\begin{align}\label{22}
&d\tau\overset{\Sigma}=e^{2\gamma-2\upsilon}\sqrt{1-\left(\frac{d\rho}{d\nu}\right)^2}d\nu=Adt,\\\label{23}
&C\overset{\Sigma}=e^{\upsilon},\quad e^{\upsilon}=\frac{1}{\rho},
\end{align}
with $1-\left(\frac{d\rho}{d\nu}\right)^2>0$. The second fundamental
form gives us the following set of equations
\begin{align}\label{24}
&e^{2\gamma-2\upsilon}[\nu_{\tau\tau}\rho_{\tau}-\rho_{\tau\tau}\nu_{\tau}
-\{\rho_\tau(\upsilon_\nu-\gamma_\nu)\}(\rho_\tau^2-\nu_\tau^2)+\nu_\tau(\gamma_\rho-\upsilon_\rho)]
\overset{\Delta}=-\frac{A'}{A^2},\\\label{25}
&e^{2\upsilon}(\nu_\tau\upsilon_\rho+\rho_\tau\upsilon_{\nu})\overset{\Delta}=\frac{BB'}{A},\quad
e^{-2\upsilon}\rho^2\left(\nu_\tau\upsilon_\rho+\rho_\tau\upsilon_\nu-\frac{\nu_\tau}{\rho}\right)\overset{\Delta}=-\frac{CC'}{A}.
\end{align}
Using Eqs.(\ref{22})-(\ref{25}) and field equations, we get
\begin{align}\label{26}
&\overset{~~\textrm{eff}}{P_r}\overset{\Sigma}=\frac{A(BC)_{,0}}{(BC)'}\overset{~~\textrm{eff}}{q}.
\end{align}
Further, we have
\begin{align}\label{27}
R|_-^+&=0,\quad f_{,RR}[\partial_\nu R|_-^+=0,\quad f_{,RR}\neq0.
\end{align}
The constraint mentioned in Eq.(\ref{26}) is due to Darmois junction
conditions. Equation (\ref{26}) shows that radial pressure on the
boundary surface is non-zero and heat flux depends along with the
dark source terms coming from $f(R,T,Q)$ gravity. Equation
(\ref{27}) is required to be obeyed over the hypersurface due to
modified gravity which affirms the continuity of Ricci curvature
invariant over $\Sigma$ even for matter thin shells.

\subsection{Oscillations}

For stability analysis of cylindrical stellar objects, we discuss
perturbed field, dynamical equations and kinematical quantities in
this section. Now, we use the linear perturbation scheme with
parameter $\epsilon$ which is non zero and very small number. We
would neglect its second and higher powers in our calculations.
Initially, the cylindrical stellar system is in hydrostatic
equilibrium but as the time passes, it is evolved and is subjected
to enter in the oscillatory phase. All the metric variables and
fluid parameters can be perturbed with the help of the following scheme \cite{15, chananiso1, chananiso2, chanrad}
\begin{align}\nonumber
&\mu(t,r)=\mu_o(r)+\epsilon{\bar{\mu}}(t,r),\quad
~~~~A(t,r)=A_o(r)+{\epsilon}\omega(t)a(r),\\\nonumber
&P_\phi(t,r)=P_{\phi o}(r)+\epsilon{\bar{P_\phi}}(t,r),\quad
B(t,r)=B_o(r)+{\epsilon}\omega(t)b(r),\\\nonumber
&P_r(t,r)=P_{ro}(r)+\epsilon{\bar{P_r}}(t,r),\quad
C(t,r)=C_o(r)+{\epsilon}\omega(t)c(r),\\\nonumber
&P_z(t,r)=P_{zo}(r)+\epsilon{\bar{P_z}}(t,r),\quad
R(t,r)=R_o(r)+{\epsilon}\omega(t)d(r),\\\nonumber &
Q(t,r)=Q_o(r)+{\epsilon}\omega(t)g(r)\\\label{28}
\end{align}
where $0<\epsilon\ll1$. Our aim is to explore unstable ranges for
cosmic anisotropic stellar filaments, therefore we confine our
investigations under which $a(r),~b(r)$, $c(r),~d(r)$ and $g(r)$ are
configuring to give frequency of oscillations to be definite
positive on the hypersurface, $\Sigma$. Applying perturbation
technique on Eqs.(\ref{12})-(\ref{17}), we get a second order
partial differential equation whose solution is of the form
\begin{align}\label{29}
\omega=\omega(t)=-\exp({\xi}t).
\end{align}
The frequency of respective dynamical system in cylindrical
configuration is calculated as
\begin{align}\nonumber
\xi^2=-\left(\frac{a}{A_o}+\frac{b}{B_o}+\frac{c}{C_o}\right)^{-1}
\left[\frac{a''}{A_o}-\frac{aA_o''}{A_o}+\frac{B_o'}{B_o}\left(\frac{c}{C_o}\right)'
+\frac{C_o'}{C_o}\left(\frac{b}{B_o}\right)'\right].
\end{align}
The perturbed profile of $f(R,T,Q)$ model is
\begin{align}\label{30}
f=R_o(1+\alpha Q_o)+\epsilon \omega(t)[d+\alpha(dQ_o+R_og)],
\end{align}
where
\begin{align}\nonumber
R_o=-\frac{2}{A_o}\left[\frac{B_o''}{B_o}+\frac{A_o''}{A_o}
+\frac{C_o''}{C_o}+\frac{B_o'C_o'}{B_oC_o}-\frac{A_o'^2}{A_o}\right]
\end{align}
and $d=d(r),~g=g(r)$. The static forms of $f(R,T,Q)$ field equations
by using perturbation scheme are
\begin{align}\nonumber
\overset{~~\textrm{eff}}{\mu_{o}}&=\frac{1}{1+\alpha
(Q_o+R_o\mu_o)}\left[\mu_o'\chi_{3o}+\mu_o\chi_{1o}+P_{ro}\chi_{4o}
+P_{\phi o}\chi_{6o}+P_{zo}\chi_{5o}+\frac{\alpha}{A_o^2}
\right.\\\nonumber &\left.\times(Q_o''+R_o'P_{r0}')-\alpha
Q_o'\psi_{2o}+\frac{\alpha R_o}{2A_o^2}
\left(\mu_o''+P_{ro}''+P_{zo}'\frac{B'_o}{B_o}+P_{\phi
o}'\frac{C'_o}{C_o}-5P_{ro}'\right.\right.\\\label{31}
&\left.\left.\times\frac{A'_o}{A_o}\right)-\frac{\beta}{2}Q_o\right],\\\nonumber
\overset{~~\textrm{eff}}{P_{ro}}&=\frac{1}{1+\alpha
(Q_o+R_o\mu_o)}\left[P_{ro}'\chi_{12o}+\mu_o\chi_{8o}+P_{ro}\chi_{7o}
+P_{\phi o}\chi_{10o}+P_{zo}\chi_{9o}-\alpha \right.\\\label{32}
&\left.\times Q_o'\psi_{2o}+\frac{\alpha R_o}{2A_o^2}
\left(\mu_{o}'\frac{A'_o}{2A_o}-P_{zo}'\frac{B'_o}{B_o}-P_{\phi
o}'\frac{C'_o}{C_o}\right)+\frac{\beta}{2}Q_o\right],\\\nonumber
\overset{~~\textrm{eff}}{P_{zo}}&=\frac{1}{1+\alpha
(Q_o+R_o\mu_o)}\left[P_{ro}\chi_{15o}+\mu_o\chi_{13o}+P_{zo}\chi_{16o}
+P_{zo}'\chi_{17o}+\chi_{14o}P_{\phi o}+\psi_{3o}\right.\\\label{33}
&\left.+\frac{\alpha R_o}{2A_o^2}
\left(\mu_{o}'\frac{A'_o}{A_o}+5P_{ro}'\frac{A_o'}{A_o}-2P_{ro}'\frac{R_o'}{R_o}-P_{ro}''+P_{zo}''
-P_{\phi o}'\frac{C'_o}{C_o}\right)\right],\\\nonumber
\overset{~~\textrm{eff}}{P_{\phi o}}&=\frac{1}{1+\alpha
(Q_o+R_o\mu_o)}\left[\chi_{13o}\mu_o+P_{ro}\chi_{15o}+\chi_{20o}P_{zo}
+\chi_{23o}P_{\phi o}'+P_{\phi
o}\chi_{21o}+\psi_{4o}\right.\\\label{34} &\left.+\frac{\alpha
R_o}{2A_o^2}
\left(\mu_{o}'\frac{A'_o}{A_o}+5P_{ro}'\frac{A_o'}{A_o}-2P_{ro}'\frac{R_o'}{R_o}-P_{ro}''+P_{\phi
o}'' -P_{z o}'\frac{B'_o}{B_o}\right)\right],
\end{align}
whereas the perturbed profile of these FEs is
\begin{align}\nonumber
\overset{\textrm{eff}}{\bar{\mu}}&=\frac{1}{1+\alpha(Q_o+R_o\mu_o)}\left[\bar{\mu}\chi_{1o}
+\dot{\bar{\mu}}\chi_{2o}+\chi_{3o}\bar{\mu}'+\chi_{4o}\bar{P_r}'
+\chi_{5o}\bar{P_z}+\chi_{6o}\bar{P_\phi}'\right.\\\nonumber
&\left.+\omega(\mu_ox_1+x_3\mu_o'+x_4P_{ro}+P_{zo}x_5+x_6P_{\phi
o})+\frac{\alpha
R_o}{2A_o^2}\left(\bar{\mu}''+\ddot{\bar{\mu}}+\bar{P_r}''+2\bar{P_r}'\right.\right.\\\nonumber
&\left.\left.\times\frac{R'_o}{R_o}
-5\bar{P_r}'\frac{A'_o}{A_o}-\bar{P_z}\frac{B_o'}{B_o}+P_{\phi}'\frac{C_o'}{C_o}\right)
+\frac{\alpha
\omega}{2A_o^2}\left(\alpha\mu_o''+dR_o''+2d'P_{ro}'-5d\right.\right.\\\nonumber
&\left.\left.P_{ro}'\frac{A'_o}{A_o}+dP_{zo}'\frac{B'_o}{B_o}
+dP_{\phi
o}'\frac{C'_o}{C_o}+2g''-4Q_o''\right)+\alpha\omega(g'\psi_{2o}-y_2Q_o')
+\alpha\omega\right.\\\nonumber &\left.\times
\frac{R_o}{2A_o^2}\left\{\frac{2a\mu_o''}{A_o}-\frac{2aP_{ro}''}{A_o}-4a\frac{P_{ro}'R'_o}{A_oR_o}
+5\frac{P_{ro}'a'}{A_o}-15a\frac{P_{ro}'R'_o}{A_o^2}-b\frac{P_{zo}'B'_o}{B_o}\right.\right.\\\nonumber
&\left.\left.
+2a\frac{P_{zo}'B'_o}{A_oB_o}-b'\frac{P_{ro}'}{B_o}+2a\frac{P_{\phi
o}'C'_o}{A_oC_o}+\left(\frac{c}{C_o}\right)'P_{\phi
o}\right\}\right]-\{\alpha\omega(g+d\mu_o)+\alpha\\\label{35} &
R_o\bar{\mu}\}\frac{\overset{\textrm{eff}}{\mu_o}}{1+\alpha(Q_o+R_o\mu_o)},\\\nonumber
\overset{\textrm{eff}}{\bar{P_r}}&=\frac{1}{1+\alpha(Q_o+R_o\mu_o)}\left[\bar{P_r}\chi_{7o}+\bar{\mu}
\chi_{8o}+\bar{P_z}\chi_{9o}+\bar{P_\phi}\chi_{10o}+\dot{\bar{P_r}}\chi_{11o}+\bar{P_r}'\right.\\\nonumber
&\left.\times \chi_{12o}+\omega
\{P_{ro}x_7+\mu_ox_8+P_{zo}x_9+P_{\phi
o}x_{10}+P_{ro}'x_{12}-\alpha(y_2Q_o'+g'\psi_{2o})\}\right.\\\nonumber
&\left.+\frac{\alpha
R_o}{2A_o^2}\left(2\ddot{\omega}\frac{g}{R_o}-\ddot{\bar{\mu}}+\bar{\mu}'\frac{A_o'}{A_o}
-\ddot{\bar{P_r}}+\bar{P_z}'\frac{B_o'}{B_o}-\bar{P_\phi}'\frac{C_o'}{C_o}\right)
+\frac{\alpha
R_o}{2A_o^2}\left(d\mu_o'\frac{A_o'}{A_o}-d\right.\right.\\\nonumber
&\left.\left.\times P_{zo}'\frac{B_o'}{B_o}-dP_{\phi
o}'\frac{C_o'}{C_o}\right)+\frac{\alpha\omega
R_o}{2A_o^2}\left(\mu_o'\frac{a'}{A_o}-3\mu_o'a\frac{A_o'}{A_o^2}+bP_{zo}'\frac{B'_o}{B_o}
-2aP_{zo}'\frac{B'_o}{A_oB_o}\right.\right.\\\label{36}
&\left.\left.+P_{zo}'\frac{b}{B_o}+2aP_{\phi
o}'\frac{C'_o}{A_oC_o}-\left(\frac{c}{C_o}\right)'P_{\phi
o}'\right)\right]-\frac{\{\alpha\omega(g+d\mu_o)+\alpha
R_o\bar{\mu}\}}{1+\alpha(Q_o+R_o\mu_o)}\overset{\textrm{eff}}{P_{ro}},\\\nonumber
\overset{\textrm{eff}}{\bar{P_z}}&=\frac{1}{1+\alpha(Q_o+R_o\mu_o)}\left[\omega(\mu_ox_{13}
+P_{ro}x_{15}+P_{zo}x_{16}+P_{ro}'x_{17}+P_{\phi
o}x_{19}+y_3)\right.\\\nonumber &\left.-\frac{\alpha
R_o}{2A_o^2}\left(\ddot{\bar{\mu}}-\bar{\mu}'\frac{A_o'}{A_o}
+2\bar{P_r}'\frac{R_o'}{R_o}-5\bar{P_r}'\frac{A_o'}{A_o}+\bar{P_r}''-\bar{P_z}''
+\ddot{\bar{P_z}}+\bar{P_\phi}'\frac{C_o'}{C_o}\right)+\alpha\omega
\right.\\\nonumber &\left.\times\frac{
R_o}{2A_o^2}\left\{\frac{\mu_o'a'}{A_o}-3\mu_o'\frac{aA_o'}{A_o^2}+4aP_{ro}'\frac{R_o'}{A_oR_o}
-5P_{ro}'\frac{a'}{A_o}+15aP_{ro}'\frac{A_o'}{A_o}+2a\frac{P_{ro}''}{A_o}\right.\right.\\\nonumber
&\left.\left. -2aP_{zo}''+2aP_{\phi
o}'\frac{C_o'}{A_oC_o}-\left(\frac{c}{C_o}\right)'P_{\phi
o}'\right\}+\bar{\mu}\chi_{13o}+\dot{\bar{\mu}}\chi_{14o}+\bar{P_{r}}\chi_{15o}
+\bar{P_{z}}\chi_{16o}\right.\\\label{37}
&\left.+P_{zo}'\chi_{17o}+\bar{P_{z}}'\chi_{17o}+\dot{\bar{P_z}}
\chi_{18o}+\bar{P_{\phi}}\chi_{19o}\right]-\frac{\{\alpha\omega(g+d\mu_o)+\alpha
R_o\bar{\mu}\}}{1+\alpha(Q_o+R_o\mu_o)}\overset{\textrm{eff}}{P_{zo}},\\\nonumber
\overset{\textrm{eff}}{\bar{P_\phi}}&=\frac{1}{1+\alpha(Q_o+R_o\mu_o)}\left[\omega(\mu_ox_{13}
+P_{ro}x_{15}+P_{zo}x_{20}+P_{\phi o}x_{21}+P_{\phi
o}'x_{23}+y_{4})\right.\\\nonumber
&\left.+\bar{\mu}\chi_{13o}+\dot{\bar{\mu}}\chi_{14o}+\bar{P_r}\chi_{15o}+
\bar{P_z}\chi_{20o}+\bar{P_\phi}\chi_{21o}+\dot{\bar{P_\phi}}\chi_{22o}
+\bar{P_\phi}'\chi_{23o}+\frac{\alpha R_o}{2A_o}\right.\\\nonumber
&\left.\times\left(\bar{P_\phi}''-\ddot{\bar{\mu}}+\bar{\mu}'\frac{A_o'}{A_o}
-2\bar{P_r}'\frac{R_o'}{R_o}+5\bar{P_r}'\frac{A_o'}{A_o}-\bar{P_r}''
-\ddot{\bar{P_\phi}}+\bar{P_z}'\frac{B_o'}{B_o}\right)+\frac{\alpha\omega}{2A_o^2}\right.\\\nonumber
&\left.\times
\left(d\mu_o'\frac{A_o'}{A_o}-2d'P'_{ro}+5dP_{ro}'\frac{A_o'}{A_o}
-dP_{ro}''+dP_{\phi
o}''-dP_{zo}'\frac{B_o'}{B_o}\right)+\frac{\alpha\omega
R_o}{2A_o^2}\right.\\\nonumber
&\left.\times\left(\mu_o'\frac{a'}{A_o}-3\mu_o'a\frac{A_o'}{A_o^2}
+4aP_{ro}'\frac{R'_o}{A_oR_o}-5P_{ro}'\frac{a'}{A_o}+15aP_{ro}\frac{A_o'}{A_o^2}
+2a\frac{P_{ro}''}{A_o}\right.\right.\\\label{38}
&\left.\left.+2a\frac{P_{\phi
o}''}{A_o}+bP_{zo}'\frac{B'_o}{B_o}-2aP_{zo}'\frac{B'_o}{A_oB_o}+b\frac{P_{zo}'}{B_o}\right)\right]
-\frac{\{\alpha\omega(g+d\mu_o)+\alpha
R_o\bar{\mu}\}}{1+\alpha(Q_o+R_o\mu_o)}\overset{\textrm{eff}}{P_{\phi
o}},\\\nonumber
\overset{\textrm{eff}}{\bar{q}}&=\frac{1}{1+\alpha(Q_o+R_o\mu_o)}\left[\bar{q}
\left\{1-\frac{3}{2}\alpha R_o^2+\frac{\alpha
R_o''}{2A_o^2}+\frac{\alpha
R_o}{2A_o^2}\left(\frac{R_o'C_o'}{R_oC_o}+\frac{R_o'B_o'}{R_oB_o}\right.\right.\right.\\\nonumber
&\left.\left.\left.+
4\frac{R_o'A_o'}{R_oA_o}+2\frac{A_o'C_o'}{A_oC_o}+2\frac{A_o'B_o'}{A_oB_o}
+\frac{2A_o''}{A_o}+\frac{2A_o'^2}{A_o}\right)\right\}+\frac{\alpha
R_o}{2A_o^2}(\bar{q}''-\ddot{\bar{q}})+\frac{\alpha
R_o}{2A_o}\right.\\\label{39}
&\left.\times\bar{q}'\left\{\frac{4A_o'}{A_o^2}+\frac{B_o'}{B_o}
+\frac{C_o'}{C_o}+\frac{2R_o'}{R_oA_o}\right\}-\frac{\alpha\dot{\omega}}{A_o^2}
\left(g'-\frac{A_o'}{A_o}-\frac{aQ_o'}{A_o}\right)\right].
\end{align}
However, the non-static forms of Eqs.(\ref{19}) and (\ref{20}) are
\begin{align}\label{40}
&\overset{~~\textrm{eff}}{\dot{\bar{\mu}}}+\dot{\omega}\eta=0,\\\nonumber
&\frac{1}{A_o}\left[\overset{~~\textrm{eff}}{\bar{\dot{q}}}+\overset{~~\textrm{eff}}{\bar{P_{r}}'}+\frac{\omega
a}{A_o}\left\{\left(\overset{~~\textrm{eff}}{P_{zo}}-\overset{~~\textrm{eff}}{P_{ro}}\right)\frac{B_o'}{B_o}
+
\left(\overset{~~\textrm{eff}}{P_{\phi
o}}-\overset{~~\textrm{eff}}{P_{ro}}\right)\frac{C_o'}{C_o}+
\left(\overset{~~\textrm{eff}}{P_{ro}}+\overset{~~\textrm{eff}}{\mu_{o}}\right)\right.\right.\\\nonumber
&\left.\left.\times
\left(\frac{a'}{a}-\frac{2A_o'}{A_o}\right)\right\}
+\omega\left\{\left(\left(\overset{~~\textrm{eff}}{P_{ro}}
-\overset{~~\textrm{eff}}{P_{\phi o}}\right)
\left(\frac{c}{C_o}\right)'+\overset{~~\textrm{eff}}{P_{ro}}-\overset{~~\textrm{eff}}{P_{zo}}\right)
\left(\frac{b}{B_o}\right)'+\frac{B_o'}{B_o}\right.\right.\\\label{41}
&\left.\left.\times\left(\overset{~~\textrm{eff}}{\bar{P_{r}}}-\overset{~~\textrm{eff}}{\bar{P_{z}}}\right)
+\left(\overset{~~\textrm{eff}}{\bar{\mu}}+\overset{~~\textrm{eff}}{\bar{P_{r}}}\right)
\frac{A_o'}{A_o}-\left(\overset{~~\textrm{eff}}{\bar{P_{\phi}}}+\overset{~~\textrm{eff}}{\bar{P_{r}}}\right)
\frac{C_o'}{C_o}\right\}+\overset{~~\textrm{eff}}{P_{ro}'}\right]+\omega
Z_2=0,
\end{align}
where
\begin{align}\nonumber
&\eta=\overset{~~\textrm{eff}}{\mu_o}\left(\frac{b}{B_o}+\frac{a}{A_o}+\frac{c}{C_o}\right)
+\overset{~~\textrm{eff}}{P_{ro}}\frac{a}{A_o}+\overset{~~\textrm{eff}}{P_{zo}}\frac{b}{B_o}
+\frac{c}{C_o}\overset{~~\textrm{eff}}{P_{\phi o}}\\\nonumber
&+S\left(2\frac{A_o'}{A_o}+\frac{C_o'}{C_o}+\frac{B_o'}{B_o}\right)+ A_oZ_3,
\end{align}
and $Z_3$ is the perturbed configurations of $Z_1$. The perturbed configuration of kinematical quantities is
\begin{align}\nonumber
\bar{\Theta}&=\left(\frac{b}{B_o}+\frac{a}{A_o}+\frac{c}{C_o}\right)\frac{\dot{\omega}}{A_o},~~
\bar{\sigma_k}=\frac{\dot{\omega}}{A_o}\left(\frac{c}{C_o}-\frac{a}{A_o}\right),~~
\bar{\sigma_s}=\frac{\dot{\omega}}{A_o}\left(\frac{b}{B_o}-\frac{a}{A_o}\right).
\end{align}

\subsection{Hydrodynamical Equation}

For the stability analysis of anisotropic dissipative cylindrical
collapse, we would construct extended form of the hydrodynamic
equation and analyze its contribution at N and PN limits in this
section. Harrison-Wheeler equation of state \cite{28} has great
impact in this context which is relation between perturbed
configuration of pressure and energy density of system given as
\begin{equation}\label{42}
\bar{P_i}=\bar{\mu}\frac{P_{i0}}{\mu_0+P_{i0}}\Gamma_1,
\end{equation}
where the stability index $\Gamma_1$ refers to measure the stiffness
of fluid content. Now, we rewrite Eq.(\ref{40}) as,
\begin{align}\nonumber
\overset{~~\textrm{eff}}{\dot{\bar{\mu}}}=-\dot{\omega}\eta.
\end{align}
making the integration of the above equation with respect to
temporal component, we get
\begin{align}\label{43}
\overset{~~\textrm{eff}}{\bar{\mu}}=-\eta\omega.
\end{align}
The (01) field equation gives
\begin{align}\label{44}
\overset{~~\textrm{eff}}{\bar{q}}=S\dot{\omega}.
\end{align}
By using Eqs.(\ref{43}) and (\ref{44}) in Eq.(\ref{42}), we get
\begin{align}\label{45}
\overset{~~\textrm{eff}}{\bar{P}_{r}}=-\Gamma_1
\frac{\overset{~~\textrm{eff}}{P_{ro}}\eta\omega}{(\overset{~~\textrm{eff}}{P_{ro}}+\overset{~~\textrm{eff}}{\mu_{o}})},\quad
\overset{~~\textrm{eff}}{\bar{P}_{z}}=-\Gamma_1
\frac{\overset{~~\textrm{eff}}{P_{z
o}}\eta\omega}{(\overset{~~\textrm{eff}}{P_{zo}}+\overset{~~\textrm{eff}}{\mu_{o}})},\quad
\overset{~~\textrm{eff}}{\bar{P}_{\phi }}=-\Gamma_1
\frac{\overset{~~\textrm{eff}}{P_{\phi
o}}\eta\omega}{(\overset{~~\textrm{eff}}{P_{\phi
o}}+\overset{~~\textrm{eff}}{\mu_{o}})}.
\end{align}
By making use of Eqs.(\ref{43})-(\ref{45}), in Eqs.(\ref{41}), the
resulting hydrodynamical equation is
\begin{align}\nonumber
&\Gamma_1\left[\frac{\overset{~~\textrm{eff}}{P_{ro}}}{\overset{~~\textrm{eff}}
{\mu_{o}}+\overset{~~\textrm{eff}}{P_{ro}}}
\left\{\eta\left\{\left(\frac{B_o'}{B_o}+\frac{C'_o}{C_o}-\frac{A_o'}{A_o}\right)+
\frac{(\overset{~~\textrm{eff}}{\mu_{o}}'+\overset{~~\textrm{eff}}{P_{ro}'})}
{(\overset{~~\textrm{eff}}{\mu_{o}}+\overset{~~\textrm{eff}}{P_{ro}})}\right\}-\eta'\right\}
-\eta\right.\\\nonumber &\times\left.
\frac{\overset{~~\textrm{eff}}{P_{ro}'}}{\overset{~~\textrm{eff}}
{\mu_{o}}+\overset{~~\textrm{eff}}{P_{ro}}}+
\frac{\overset{~~\textrm{eff}}{P_{zo}}}{\overset{~~\textrm{eff}}
{\mu_{o}}+\overset{~~\textrm{eff}}{P_{zo}}}\eta\frac{B_o'}{B_o}+
\frac{\overset{~~\textrm{eff}}{P_{\phi o}}}{\overset{~~\textrm{eff}}
{\mu_{o}}+\overset{~~\textrm{eff}}{P_{\phi
o}}}\eta\frac{C_o'}{C_o}\right]+\xi^2S=-\frac{a}{A_o}\overset{~~\textrm{eff}}{P_{ro}'}+\overset{~~\textrm{eff}}{P_{ro}}
\\\nonumber
&\times\left\{\frac{aB_o'}{A_oB_o}+\frac{a'}{a}-\frac{C_o'}{C_o}-\frac{2A'_o}{A_o}
-\left(\frac{c}{C_o}\right)'-\left(\frac{b}{B_o}\right)'\right\}+\overset{~~\textrm{eff}}{P_{zo}}
\left\{\left(\frac{b}{B_o}\right)'-\frac{aB_o'}{A_oB_o}\right\}
\\\label{46} &+\overset{~~\textrm{eff}}{\mu_{o}}\left(\frac{a'}{a}-\frac{2A'_o}{A_o}\right)
+\overset{~~\textrm{eff}}{P_{\phi o}}
\left\{\left(\frac{c}{C_o}\right)'-\frac{C_o'}{C_o}\right\}
+A_oZ_2+\frac{\eta A'_o}{A_o},
\end{align}
which is the corresponding cylindrical collapse equation within the
framework of $f(R,T,Q)$ gravity. In above equation, the terms
containing adiabatic index $\Gamma_1$ treated as the generator of
pressure gradients and anti-gravitational effects while the rest
ones would generate gravity force.

\subsubsection{N Approximation}

In order to discuss the instability eras at $N$ epoch, we consider
\begin{equation*}
\mu_0\gg P_{i0}, \quad A_0=1, \quad B_0=1.
\end{equation*}
Using these conditions, the respective collapse equation (\ref{46})
turns out to be
\begin{align}\nonumber
&\left[\overset{~~\textrm{eff}}{\mu_{o}}\left(a+b+\frac{c}{C_o}\right)\right]
\Gamma_1=\overset{~~\textrm{eff}}{\mu_{o}}(a'/a)+S\xi^2_N+\Pi+Z_2,
\end{align}
where
\begin{align}\label{47}
\Pi=&\left(\overset{~~\textrm{eff}}{P_{\phi
o}}+\overset{~~\textrm{eff}}{P_{ro}}\right)\frac{C'_o}{C_o}+b'\left(\overset{~~\textrm{eff}}{P_{zo}}-\overset{~~\textrm{eff}}
{P_{ro}}\right)+\left(\frac{c}{C_o}\right)'
\left\{-\left(\overset{~~\textrm{eff}}{P_{\phi
o}}+\overset{~~\textrm{eff}}{P_{ro}}\right)\right\}+a(-\overset{~~\textrm{eff}}{P_{ro}}'),
\end{align}
includes effects induced by local anisotropic pressure and dark
sector terms coming from $f(R,T,Q)$ gravity in the instability of
the cylindrical compact objects. In $f(R,T,Q)$ gravity, the evolving
cylindrical anisotropic stellar object in account of degrees of
freedom will be in phase of hydro-static equilibrium whenever it
satisfies
\begin{align}\label{48}
&\Gamma_1=\frac{|\overset{~~\textrm{eff}}{\mu_{o}}(a'/a)+\Pi+S\xi^2_N+Z_2|}
{|\overset{~~\textrm{eff}}{\mu_{o}}\left(a+\frac{c}{C_o}+b\right)|}.
\end{align}
If upon evolution, the system is able to get the state under which
the effects of
$|\overset{~~\textrm{eff}}{\mu_{o}}(a'/a)+S\xi^2_N+\Pi+Z_2|$ and
$|\overset{~~\textrm{eff}}{\mu_{o}}\left(a+b+\frac{c}{C_o}\right)|$
are equal, then adiabatic index would be 1. Then, the system will be
at equilibrium state. However, the relation
\begin{align}\label{49}
&\Gamma_1<\frac{|\overset{~~\textrm{eff}}{\mu_{o}}(a'/a)+S\xi^2_N+\Pi+Z_2|}
{|\overset{~~\textrm{eff}}{\mu_{o}}\left(a+\frac{c}{C_o}+b\right)|}
\end{align}
shows that the given system is in unstable region. This constraint shows that
adiabatic index depends upon $\overset{~~\textrm{eff}}{\mu_{o}},~S,~\Pi$ and $Z_i's$. Now, we shall discuss
the role of these matter variables on the onset of instability regions.
\begin{itemize}
\item The quantity $\overset{~~\textrm{eff}}{\mu_{o}}$ describes the energy density of the dissipative
anisotropic matter configurations along with the additional
gravitational field due to $f(R,T,Q)$ extra curvature terms. Since
we have evaluated the above condition to describe the collapsing
scenario, therefore the static profile of energy density along with
the corresponding dark source terms will be much greater than the
static configurations of effective pressure components.
\item The term $\Pi$ constitutes pressure anisotropic effects and is given in Eq.$(\ref{47})$. One can see from expression
(\ref{49}) that $\Pi$ tends to produce increments in the instability
regions of the self-gravitating cylindrical systems, whence $\Pi>0$.
For this purpose, we need to impose some conditions, i.e.,
$\overset{~~\textrm{eff}}{P_{zo}}>\overset{~~\textrm{eff}}
{P_{ro}},~ |\overset{~~\textrm{eff}}{P_{\phi
o}}+\overset{~~\textrm{eff}}{P_{ro}}|$ and
$|\overset{~~\textrm{eff}}{P_{ro}}'|$. If
$\overset{~~\textrm{eff}}{P_{zo}}<\overset{~~\textrm{eff}}
{P_{ro}},~ |\overset{~~\textrm{eff}}{P_{\phi
o}}+\overset{~~\textrm{eff}}{P_{ro}}|$ and
$|\overset{~~\textrm{eff}}{P_{ro}}'|$, then $\Pi<0$. At that region
of cylindrical spacetime, anisotropic effects would have opposite
effects and would decrease the regions of instabilities of the
collapsing cosmic filaments. This behavior of anisotropic pressure
in spherical distributions of matter increases the instabilities of collapsing stars. The similar behavior of
anisotropic pressure has been found by Chan et al.
\cite{chananiso1,chananiso2} for the case of spherical distribution
of matter.
\item The quantity $\xi^2_NS$ comes from the (01) component
of the $f(R,T,Q)$ field equations and contributes the influences of
heat radiations in the instability constraint. This term tends to
increase the instability limits against gravitation implosion,
thereby making our system less stable. The same type of radiation
effects have been explored by Chan et al. \cite{chanrad} for the
case of spherically symmetric relativistic configurations.
\item The terms $Z_i's$ correspond to dark source terms induced from
non-minimal coupled modified gravity, i.e., $f(R,T,Q)$ gravity.
These terms are inducing DE effects in the instability constraints
and are creating anti gravity effects due to their non-attractive
nature, thereby producing stability against gravitational collapse.
The presence of these terms have completely modified the landscape
of theoretical and mathematical aspects of dynamical evolving cosmic
stellar filaments. This result is well-consistent with those
obtained for dynamical instability of spherical expansion-free
matter configurations \cite{20,dark} in modified gravity.
\end{itemize}

If the $f(R,T,Q)$ gravity forces generated by
$|\overset{~~\textrm{eff}}{\mu_{o}}(a'/a)+\Pi+S\xi^2_N+Z_2|$ are
greater than
$|\overset{~~\textrm{eff}}{\mu_{o}}\left(a+b+\frac{c}{C_o}\right)|$,
then expression (\ref{49}) will impart value of $\Gamma_1$ greater
than 1. This means that the forces of anti-gravity and principal
stresses along with radiations would provide $\Gamma_1>1$ dynamical
stable condition.

\subsubsection{pN Approximation}

To gain pN instability constraints, we take
$A_0(r)=1-\phi,~B_0(r)=1+\phi$, with relativistic corrections up to
the first order in  $\phi(r)=\frac{m_0}{r}$. Using
hydrodynamical equation (\ref{46}), the value of $\Gamma_1$
can be recasted as
\begin{align}\label{50}
\Gamma_1=\frac{E_{pN}}{F_{pN}},
\end{align}
where
\begin{align}\nonumber
E_{pN}&=\frac{\overset{~~\textrm{eff}}{P_{ro}}}{\overset{~~\textrm{eff}}
{\mu_{o}}+\overset{~~\textrm{eff}}{P_{ro}}}\left[-\eta'_{pN}+\eta_{pN}
\left\{\frac{\overset{~~\textrm{eff}}{\mu_{o}'}}{\overset{~~\textrm{eff}}
{\mu_{o}}+\overset{~~\textrm{eff}}{P_{ro}}}+\frac{C'_o}{C_o}+\phi'\frac{\overset{~~\textrm{eff}}{P_{zo}}}
{\overset{~~\textrm{eff}}
{\mu_{o}}+\overset{~~\textrm{eff}}{P_{zo}}}(1-\phi)+2\phi'\right.\right.\\\nonumber
&\left.\left.+
\frac{\overset{~~\textrm{eff}}{P_{\phi o}}}{\overset{~~\textrm{eff}}
{\mu_{o}}+\overset{~~\textrm{eff}}{P_{\phi
o}}}\frac{C'_o}{C_o}\right\}\right],\\\nonumber
F_{pN}&=-a(1+\phi)\overset{~~\textrm{eff}}{P_{ro}}'
+(\overset{~~\textrm{eff}}{P_{ro}}-\overset{~~\textrm{eff}}{P_{\phi
o}})S_2+
(\overset{~~\textrm{eff}}{P_{ro}}-\overset{~~\textrm{eff}}{P_{zo}})S_1+
(\overset{~~\textrm{eff}}{\mu_{o}}+\overset{~~\textrm{eff}}{P_{ro}})S_3,\\\nonumber
&-\eta_{pN}(1+\phi)\phi'+Z_2(1-\phi)+\xi^2_{pN}S.
\end{align}
It can be seen from the above that $E_{pN}$ contains gravity effects due to components of effective anisotropic
pressure, while $F_{pN}$ contains effects due to effective heat dissipation, pressure anisotropy and dark source terms coming from
non-minimal matter geometry couplings. In order to calculate instability regions, we assume that each term in $E_{pN}$ and
$F_{pN}$ are positive. The cylindrically symmetric anisotropic matter dissipative matter distributions
will be in the state of equilibrium, if it satisfies Eq.(\ref{50}). Furthermore, if
during evolution $F_{pN}$ attains a value equal to $E_{pN}$, then the condition of
equilibrium at pN approximations will be $$\Gamma_1=1.$$ The instability regimes can be found from collapse equation
as
\begin{align}\label{51}
\Gamma_1<\frac{E_{pN}}{F_{pN}},
\end{align}
This equation shows that the unstable regime of our system depends
upon the contribution of anti-gravity effects, adiabatic index,
$\Gamma_1$ and principal stresses variation along with $f(R,T,Q)$
gravity. The system will be more stable due to $\alpha RQ$ factor
as compared to GR, thereby favoring the results of \cite{dark}. However, in the cylindrical
system, the relativistic corrections because of heat
flow lowers the unstable epochs of $\Gamma_1$. Thus, the effects of
radiations is qualitatively same as in the case of spherical star. If the contributions
of $E_{pN}$ is greater than $F_{pN}$, then expression (\ref{50}) provides the following instability constraint
$$\Gamma_1<1.$$
Since, we did not consider expansion-free condition, therefore all of our results of N and pN approximations
depends upon the fluid stiffness parameter, i.e., adiabatic index. This suggests the importance of $\Gamma_1$
coming in the Harrison Wheeler equation of state. Our this result is well-consistent with those obtained for
spherical stars \cite{15, chananiso1, chananiso2, chanrad}.

\section{Concluding Remarks}

This paper is devoted to study the dynamical stability of
cylindrical stellar model in the frame of $f(R,T,Q)$ gravity. The
system is made of anisotropic matter distribution along with
dissipation in terms of heat flow. We have explored the FEs for
cylindrical symmetric space-time in this theory. Also, the dynamical
equation are formulated by using the contraction of Bianchi
identities. By imposing the perturbation scheme on geometry as well
as material functions, the static and perturbed profile of the
field, kinematical and dynamical equations have designed.

In GR, the dynamical instability of anisotropic star depends upon
the adiabatic index $\Gamma_1$ having numerical value $\frac{4}{3}$
\cite{14}. The internal pressure of star being strong enough in
contrast to the weight of layers that makes the star stable, in that
case $\Gamma_1>\frac{4}{3}$. When $\Gamma_1<\frac{4}{3}$, the size
of layers become larger than the inside pressure of star
consequently dynamical instability occurs due to collapse. Here, we
have examined the the role of dark source terms coming from
$f(R,T,Q)$ gravity and heat radiation on the instability regimes of
the respective cylindrical system. Initially, we have assumed that
our system is in state of hydro-static equilibrium and facing
oscillations as the time passes. Hence, the forthcoming equations
are applied to construct the modified form of hydrodynamical
equation which is further examined with N and pN approximations. In
this framework, adiabatic index supported by equation of state has
been used to measure the stiffness of matter composition.

We have considered a well-consistent $f(R,T,Q)$ model and checked
its impact in the dynamical evolution of locally anisotropic and
dissipative cylindrical system. It is worthwhile noticed that the
extra curvature terms appearing due to the modification in gravity,
tend to move the system into the stable window due to their
non-attractive nature. The effects of heat flux are represented in
the variable $S$ in our paper. From (\ref{49}), it is noted that
evolving system would undergoes in a more unstable phase at N regime
due to effects of dissipation induced by heat flux. The relativistic
correction due to heat flow can be observed from Eq. (\ref{50})
which is expressed with the help of $\Gamma_1$ in $f(R,T,Q)$
gravity. This reveals that various physical characteristics of
matter has a remarkable relevance in structure formation and
development of self-gravitating system. Therefore, we conclude that
the system will be in hydro-equilibrium state at N epoch, if it
holds up its range as specified in (\ref{48}).

\renewcommand{\theequation}{A\arabic{equation}}
\setcounter{equation}{0}
\section*{Appendix A}

The quantities $\chi_i$'s used in the field equations are
\begin{align}\nonumber
\chi_1&=1+f_T+f_Q\left[\frac{-3R}{2}+\frac{1}{A^2}\left\{\frac{4\dot{A}^2}{A^2}-2\frac{\ddot{A}}{A}+\frac{A'^2}{A^2}-\frac{A'^2}{A}
-\frac{\dot{A}^2}{A}+\frac{A''}{A}-\frac{\dot{A}}{A}\right.\right.\\\nonumber
&\left.\left.\times\left(\frac{\dot{B}}{B}+\frac{\dot{C}}{C}\right)+\frac{{A'}}{A}\left(\frac{{B'}}{B}+\frac{{C'}}{C}\right)\right\}\right]
+\frac{\dot{f_Q}}{A^2}\left(\frac{3\dot{A}}{2A}-\frac{\dot{B}}{2B}-\frac{\dot{C}}{2C}\right)+\frac{f_Q''}{2A^2}\\\nonumber
&+\frac{f_Q'}{2A^2}
\left(\frac{B'}{B}+\frac{C'}{C}\right),\\\nonumber
\chi_2&=\frac{f_Q}{A^2}\left(\frac{\dot{B}}{B}-\frac{9\dot{A}}{2A}-\frac{\dot{C}}{C}\right),\quad
\chi_3=\frac{1}{A^2}\left[f_Q''
+\frac{f_Q'}{2}\left(\frac{B'}{B}+\frac{C'}{C}\right)\right],\\\nonumber
\chi_4&=\frac{1}{A^2}\left[f''_Q+f_Q\left(\frac{4A'^2}{A^2}-\frac{A''}{A}-\frac{\dot{A}^2}{A^2}\right)-\frac{5A'}{A}f_Q'
-\frac{\dot{A}}{2A}\dot{f_Q}\right],\\\nonumber
\chi_5&=\frac{1}{A^2}\left[f_Q\left(\frac{\dot{B}^2}{B^2}-\frac{B'^2}{B^2}\right)-\frac{\dot{B}}{2B}\dot{f_Q}
+\frac{B'}{2B}f'_Q\right],\\\nonumber
\chi_6&=\frac{1}{A^2}\left[f_Q\left(\frac{\dot{C}^2}{C^2}-\frac{C'^2}{C^2}\right)-\frac{\dot{C}}{2C}\dot{f_Q}
+\frac{C'}{2C}f'_Q\right],\\\nonumber
\chi_7&=1+f_T-\frac{3R}{2}f_Q+\frac{f_Q}{A^2}\left[\frac{2A''}{A}-\frac{3A'^2}{A^2}-2\frac{\dot{A}^2}{A^2}
-\frac{\ddot{A}}{A}-\frac{\dot{A}}{A}\left(\frac{\dot{B}}{B}+\frac{\dot{C}}{C}\right)\right.\\\nonumber
&\left.
+\frac{A'}{A}\left(\frac{B'}{B}+\frac{C'}{C}\right)\right]+\frac{f_Q'}{A^2}\left(\frac{7A'}{A}
+\frac{B'}{2B}+\frac{C'}{2C}\right)-\frac{\dot{f_Q}}{2A^2}\left(\frac{3\dot{A}}{A}
+\frac{2\dot{B}}{B}\right.\\\nonumber
&\left.+\frac{\dot{C}}{2C}\right)-\frac{\ddot{f_Q}}{2A^2},\quad
\chi_{11}=\frac{-f_Q}{2A^2}\left(3\frac{\dot{A}}{A}+\frac{\dot{B}}{B}+\frac{\dot{C}}{C}\right)-\frac{\dot{f_Q}}{A^2}\\\nonumber
\chi_9&=\frac{f_Q}{A^2}\left(\frac{\ddot{A}}{A}-4\frac{\dot{A}^2}{A^2}-\frac{A'^2}{A^2}\right)
+\frac{1}{2A^2}\left(\frac{5\dot{A}}{A}\dot{f_Q}-\ddot{f_Q}+\frac{A'}{A}f_Q'\right),\\\nonumber
\chi_{10}&=\frac{f_Q}{A^2}\left(\frac{{C'^2}}{C^2}-\frac{\dot{C}^2}{C^2}\right)
+\frac{1}{2A^2}\left(\frac{\dot{A}}{A}\dot{f_Q}-\frac{C'}{C}f_Q'\right),\\\nonumber
\chi_{12}&=\frac{f_Q}{2A^2}\left(\frac{9{A'}}{A}+\frac{{B'}}{B}+\frac{{C'}}{C}\right)-\frac{2{f_Q'}}{A^2},\\\nonumber
\chi_{13}&=\frac{f_Q}{A^2}\left(\frac{\ddot{A}}{A}-\frac{4\dot{A}^2}{A^2}-\frac{A'^2}{A^2}\right)+\frac{1}{2A^2}
\left(\frac{5\dot{A}}{A}\dot{f_Q}+\frac{A'}{A}f_Q'-\ddot{f_Q}\right),\\\nonumber
\chi_{14}&=\frac{1}{2A^2}\left(\frac{5\dot{A}}{A}f_Q-2\dot{f_Q}\right),\\\nonumber
\chi_{15}&=\frac{f_Q}{A^2}\left(\frac{A''}{A}-\frac{4A'^2}{A^2}-\frac{\dot{A}^2}{A^2}\right)
+\frac{1}{A^2}\left(\frac{5A'}{A}f'_Q+\frac{\dot{A}}{2A}\dot{f_Q}-\frac{f''_Q}{2}\right),\\\nonumber
\chi_{16}&=\frac{f_Q}{A^2}\left\{\frac{B''}{B}-\frac{\ddot{B}}{2B}-\frac{\dot{B}}{B}\left(\frac{\dot{C}}{C}+\frac{4\dot{B}}{B}\right)
+\frac{B'}{B}\left(\frac{C'}{C}+\frac{3B'}{B}\right)-\frac{\dot{f_Q}}{A^2}\left(\frac{4\dot{B}}{B}\right.\right.\\\nonumber
&\left.\left.+\frac{\dot{C}}{2C}\right)\right\}+1+f_T-\frac{3Rf_Q}{2}+\frac{1}{2A^2}(f_Q''-\ddot{f_Q}),\\\nonumber
\chi_{17}&=\frac{f_Q}{A^2}\left(\frac{4B'}{B}+\frac{C'}{C}\right)+\frac{f'_Q}{A^2},~\chi_{18}=\frac{f_Q}{A^2}
\left(\frac{2\dot{B}}{B}+\frac{\dot{C}}{2C}\right)+\frac{\dot{f_Q}}{A^2}\\\nonumber
\chi_{19}&=\frac{f_Q}{A^2}
\left(\frac{C'^2}{C^2}-\frac{\dot{C}^2}{C^2}\right)+\frac{1}{2A^2}\left(\frac{\dot{C}}{C}\dot{f_Q}-\frac{C'}{C}f'_Q\right),\\\nonumber
\chi_{20}&=\frac{f_Q}{A^2}\left(\frac{B'^2}{B^2}-\frac{\dot{B}^2}{B^2}\right)+\frac{1}{2A^2}\left(\frac{\dot{B}}{B}\dot{f_Q}
-\frac{B'}{B}f_Q'\right),\\\nonumber
\chi_{21}&=\frac{f_Q}{A^2}\left\{\frac{C''}{C}+\frac{C'}{C}\left(\frac{B'}{B}+\frac{C'}{C}\right)
-\frac{\dot{C}}{C}\left(\frac{\dot{B}}{B}+4\frac{\dot{C}}{C}\right)-\frac{\ddot{C}}{C}\right\}
+\frac{f_Q'}{2A^2} \left(\frac{4C'}{C}\right.\\\nonumber
&\left.+\frac{B'}{B}\right)-\frac{\dot{f_Q}}{2A^2}\left(\frac{\dot{B}}{B}+\frac{5\dot{C}}{C}\right)
+\frac{1}{2A^2}(f_Q''-\ddot{f_Q})+1+f_T-\frac{3R}{2}f_Q,\\\nonumber
\chi_{22}&=-\frac{\dot{f_Q}}{A^2}-\frac{f_Q}{2A^2}\left(\frac{\dot{B}}{B}+\frac{5\dot{C}}{C}\right),~
\chi_{23}=-\frac{f_Q'}{A^2}+\frac{f_Q}{2A^2}\left(\frac{B'}{B}+\frac{5C'}{C}\right).
\end{align}
The expressions of $\psi_i$'s are
\begin{align}\nonumber
\psi_1&=\frac{1}{A^2}\left(\frac{\dot{A}}{A}+\frac{\dot{B}}{B}+\frac{\dot{C}}{C}\right),\\\nonumber
\psi_2&=\frac{1}{A^2}\left(\frac{{A'}}{A}-\frac{{B'}}{B}-\frac{{C'}}{C}\right),\\\nonumber
\psi_3&=\frac{1}{A^2}\left(\ddot{f_R}-f_R''+\frac{\dot{C}}{C}\dot{f_R}+\frac{C'}{C}f'_R\right),\\\nonumber
\psi_4&=\frac{1}{A^2}\left(\ddot{f_R}-f_R''+\frac{\dot{B}}{B}\dot{f_R}+\frac{B'}{B}f'_R\right).
\end{align}
The expressions for $S_i$'s are mentioned as follows
\begin{align}\nonumber
S_1&=a\phi'-[b(1-\phi)]',~S_2=\frac{C'_o}{C_o}-\left(\frac{c}{C_o}\right)',~S_3=\frac{a'}{a}+2\phi'(1+\phi).
\end{align}
\end{document}